\documentclass[12pt]{article}

\usepackage{amsmath,amssymb}
\usepackage{amsthm}
\usepackage{graphicx}
\numberwithin{equation}{section}
\usepackage{cite}
\usepackage{ascmac}
\usepackage[breaklinks=true]{hyperref}
\usepackage{mathrsfs}
\newcommand{\er}[1]{Eq.~\eqref{#1}}
\newcommand{\ers}[1]{Eqs.~\eqref{#1}}
\newcommand{\qtq}[1]{\quad\text{#1}\quad}

\newcommand{\hc}{\text{h.c.}}
\newcommand{\bb}{\mathbb}

\renewcommand{\b}{\bar}
\renewcommand{\d}{\dot}
\newcommand{\pd}[2]{\frac{\partial{#1}}{\partial{#2}}}

\newcommand{\sr}{\sqrt}
\newcommand{\bs}{\boldsymbol}

\newcommand{\fr}{\frac}

\newcommand{\der}{\partial}
\newcommand{\lb}{\left}
\newcommand{\rb}{\right}
\renewcommand{\(}{\left(}
\renewcommand{\)}{\right)}

\newcommand{\bmx}{\left(\begin{matrix}}
\newcommand{\emx}{\end{matrix}\right)}

\usepackage[top=2.828cm,bottom=2.828cm, left=3cm,right=3cm]{geometry}
\begin{document}
\allowdisplaybreaks[4]
\begin{titlepage}
\hfill 
\vspace{-1em}
\def\thefootnote{\fnsymbol{footnote}}%
   \def\@makefnmark{\hbox
       to\z@{$\m@th^{\@thefnmark}$\hss}}%
 \vspace{3em}
 \begin{center}%
  {\Large 
Dual formulations of vortex strings \\
in 
supersymmetric Abelian Higgs model
  \par
   }%
 \vspace{1.5em} 
  {\large
Muneto Nitta\footnote{nitta@phys-h.keio.ac.jp}
and
Ryo Yokokura\footnote{ryokokur@keio.jp}
   \par
   }%
 \vspace{1.5em}
{\small \it
Department of Physics \& 
Research and Education Center for Natural Sciences,\\
Keio University, Hiyoshi 4-1-1, Yokohama, Kanagawa 223-8521, Japan
 \par
}
  \vspace{1em} 
   {\large 
    }%
 \end{center}%
 \par
\vspace{1.5em}%
\begin{abstract}
We discuss dual formulations of vortex strings 
(magnetic flux tubes) in the four-dimensional ${\cal N} =1$ 
supersymmetric Abelian Higgs model
with the Fayet--Iliopoulos term
in the superspace formalism.
The Lagrangian of the model is dualized into a Lagrangian of 
the $BF$-type described by 
a chiral spinor gauge superfield including a 2-form gauge 
field.
The dual Lagrangian is further dualized into a Lagrangian given by a chiral spinor superfield including a massive 2-form field.
In both of the dual formulations,
we obtain a superfield into which the vortex strings 
and their superpartners are embedded.
We show the dual Lagrangians in terms of a superspace and a component formalism.
In these dual Lagrangians, 
we explicitly show that 
the vortex strings of the original model are 
described by a string current electrically coupled with the 2-form gauge field
or the massive 2-form field.
\end{abstract}
\end{titlepage}
 \setcounter{footnote}{0}%
\def\thefootnote{$*$\arabic{footnote}}%
   \def\@makefnmark{\hbox
       to\z@{$\m@th^{\@thefnmark}$\hss}}%
\tableofcontents
\section{Introduction}

To understand phases of gauge theories is one of the important issues in quantum field theories.
In the gauge theories, there is the so-called Higgs phase 
where a gauge field becomes massive.
One of the simplest renormalizable 
theories describing the Higgs phase may be the Abelian Higgs model,
where a $U(1)$ gauge field is coupled with 
a complex scalar field charged under the $U(1)$ symmetry.
In the Higgs phase of the Abelian Higgs model, 
the $U(1)$ gauge field eats a phase part of the complex scalar field, 
and becomes massive. 
Furthermore, there can exist  
extended objects of 
spatial dimension one 
as solutions of the equation of motion (EOM)
in the Higgs phase.
The extended objects are so-called 
Abrikosov--Nielsen--Olesen (ANO)
vortex strings~\cite{Abrikosov:1956sx,Nielsen:1973cs}.
The ANO vortex strings are magnetic flux tubes 
which have topological charges,
and they can be regarded as topological 
solitons.
Such vortex strings arise in many contexts such as 
type-II superconductors~\cite{Abrikosov:1956sx} 
in condensed matter physics 
as well as
cosmic strings~\cite{Vilenkin:1981zs,Vilenkin:1981iu,Vilenkin:1981kz} 
in cosmology (see e.g., Refs.~\cite{Vilenkin:1984ib,Vilenkin:2000jqa} as a review).

While the ANO vortex strings are introduced as 
solutions to the EOM, they can be seen as 
charged objects associated with gauge fields by using 
dual transformations.
For the Abelian Higgs model, 
there are at least two dual formulations.
One is to dualize the phase of the scalar field to a 
2-form gauge 
field~\cite{Hayashi:1973fst,Cremmer:1973mg,Kalb:1974yc,Nambu:1974zg}.
In this dual formulation, the original 1-form gauge field and the dualized 2-form gauge field are massive by the topological coupling (the so-called $BF$ coupling) between them~\cite{Horowitz:1989ng,Allen:1990gb}.
In this dual formulation, the ANO vortex strings are described by  
a conserved string current which is electrically coupled with
the 2-form gauge field~\cite{Henneaux:1986ht,Lee:1993ty}.
Another is to dualize the massive 1-form field to
a massive 2-form field~\cite{Takahashi:1970ev}.
The 1-form gauge field becomes massive after eating 
the phase of the complex scalar field.
The dual 2-form field can be regarded as
 a 2-form gauge field eating 
the 1-form gauge field by a St\"uckelberg coupling.
In this dual formulation too, the ANO vortex strings are 
dualized to a string current electrically coupled with the massive 
2-form field~\cite{Sugamoto:1978ft,Kawai:1980qq}.
The dual transformation was applied to a
finite temperature phase transition of the Abelian Higgs model~\cite{Ramos:2005yy}.

In the Abelian Higgs model,
the positions of the ANO vortex strings are characterized by 
zero points of the complex scalar field,
and the dual string current
are described by the singularities due to 
a multivalued part of the phase of the complex scalar field around the zero points~(see e.g., Ref.~\cite{Kleinert:2008zzb}).
The dual formulation with ANO vortex strings can be obtained 
by splitting the complex scalar field into the regular part 
and the singular part.
The phase in the regular part of the complex scalar field 
can be dualized into the 2-form gauge field.
On the other hand, the phase of the singular part, which is the multivalued function, is dualized to the string current.

There are some virtues of the dual transformations.
One virtue of the dual formulations is that 
the topological charge of the ANO vortex strings can be 
simply understood as the conserved charge 
associated with the gauge symmetry for 
the 2-form field~\cite{Davis:1989zf}.
Another virtue is that 
the ANO vortex strings become fundamental degrees of freedom 
in contrast to the original theory.

In the literature,
there are some generalizations of the duality of ANO vortex strings 
in the Abelian Higgs model.
One is the case of global strings in the Goldstone model, that is, a $U(1)$ Higgs model without a gauge interaction.
In this case, a Nambu--Goldstone boson 
associated with the spontaneously broken global $U(1)$ symmetry
is dualized to a massless 2-form field, 
and global strings are electrically coupled to the 2-form field 
\cite{Davis:1989gn,Lee:1993ty}.
These strings are axion strings in cosmology 
and superfluid vortices in superfluids in condensed matter physics.
Another generalization is the case of non-Abelian gauge theories.
An $SU(2)$ gauge theory coupled with 
one complex (two real) adjoint Higgs fields 
are known to admit ${\mathbb Z}_2$ strings \cite{Nielsen:1973cs}.
A non-Abelian duality in this case was obtained in Ref.~\cite{Seo:1979id}, where the dual Lagrangian is described by 
a non-Abelian 2-form field \cite{Seo:1979id,Freedman:1980us} 
coupled with ${\mathbb Z}_2$ strings.
Another case is an $SU(3)$ gauge theory coupled 
with three by three complex Higgs fields in the fundamental representation, 
relevant for QCD at high density and low temperature.
This theory admits a non-Abelian vortex (color flux tubes) 
\cite{Balachandran:2005ev,Eto:2009kg,Eto:2013hoa}, 
accompanied with non-Abelian ${\mathbb C}P^2$ moduli 
\cite{Nakano:2007dr},
and a non-Abelian duality of non-Abelian vortices in this theory was obtained in Refs.~\cite{Hirono:2010gq,Eto:2013hoa}.

In general, there are attractive and repulsive forces among
 the ANO vortex strings intermediated by Higgs and 
 gauge fields, respectively.
For type-II (I) superconductors, the gauge field is lighter (heavier) than the Higgs field, thereby repulsion (attraction) is dominant.
The multiple 
vortex strings become stable if the two forces are balanced 
at the critical coupling between type-I and type-II superconductors.
Such a state is called 
a Bogomol'nyi--Prasad--Sommerfield (BPS) 
state~\cite{Bogomolny:1975de,Prasad:1975kr}.
In the BPS state, the total mass of the ANO vortex strings is proportional to the total topological charge 
(see e.g., Ref.~\cite{Shifman:2009zz}). 

Supersymmetry (SUSY) gives us non-perturbative aspects of 
BPS states~\cite{deVega:1976xbp}.
The BPS states preserve half of the SUSY charges if the 
theories are embedded into SUSY theories.
Since the BPS states are protected by SUSY, 
the BPS states are stable against quantum corrections 
\cite{Witten:1978mh}.
The SUSY Abelian Higgs model~\cite{Fayet:1975yh} can be constructed 
by using a vector gauge superfield (so-called 1-form prepotential)
with a Fayet--Iliopoulos (FI) term~\cite{Fayet:1974jb}
and chiral superfields charged under the $U(1)$ gauge
symmetry.
In particular, 
the ANO vortex strings can be constructed by 
using a D-term potential~\cite{Penin:1996si,Davis:1997bs,Dvali:2003zh,Copeland:2003bj,Dvali:2003zj}.

The dual formulations of the SUSY Abelian Higgs model
are possible.
In SUSY theories, the duality between 
the scalar field and the 2-form field can be extended
into the duality between a chiral superfield 
and a chiral spinor 
gauge superfield~\cite{Siegel:1979ai,Gates:1980ay,Lindstrom:1983rt}
which we will call ``2-form prepotential''~\cite{Gates:1983nr}.
This is because the 2-form gauge field can be embedded into the chiral spinor gauge superfield.
Furthermore, the duality between a massive 1-form field and 
a massive 2-form field can also be understood as 
the duality between a real superfield and a chiral spinor superfield~\cite{DAuria:2004psr,Louis:2004xi,Kuzenko:2004tn,Louis:2007nd}.
Such dual transformations 
were extended to supergravity (SUGRA)~\cite{Kuzenko:2004tn}
and extended SUSY theories~\cite{Kuzenko:2004tn}.
However, the superfield 
descriptions of the dual formulations of the ANO vortex strings 
in the SUSY context have not been understood so far.
The above mentioned dual formulations in SUSY theories 
only describe the regular part without singularities.
In order to understand non-perturbative aspects 
of the ANO vortex strings in SUSY theories,
it is plausible to dualize the SUSY Abelian Higgs model 
including the ANO vortex strings in a manifestly SUSY way.

In this paper, 
we show the dual formulations of the four-dimensional (4D) 
${\cal N} =1$ SUSY Abelian Higgs model 
including the ANO vortex strings.
We use the superspace formalism in order to give the 
manifestly SUSY theories.
There are at least two ways to dualize the Lagrangian
of the Abelian Higgs model as mentioned above.
We discuss both of the dual transformations to the theories with
a 2-form gauge field and a massive 2-form field.
In both of the dual formulations, 
we show the dual transformations of the ANO vortex strings in terms 
of superfields.
As in the bosonic Abelian Higgs model, 
we split a chiral superfield describing 
the complex scalar field into the regular part and the singular part.
For the regular part, there are no zero points of the 
complex scalar field.
Therefore, the regular part of the 
chiral superfield can be dualized 
into the 2-form prepotential.
For the singular part, the duality transformations give 
us the electrical coupling of the 2-form prepotential
with a superfield given by the singular part.
We show that the superfield given by the singular part 
has the string current as well as superpartners of the string current
by the component expression of the dual Lagrangian.
We can further dualize the 1-form prepotential.
In this dual transformation, the Lagrangian can be written
in terms of a massive chiral spinor superfield 
and the superfield into which the string current is embedded.

This paper is organized as follows.
In section~\ref{AH}, we review the dual transformations 
of ANO vortex strings in the Abelian Higgs model without SUSY.
 In section~\ref{SAH}, we show the duality transformations
of ANO vortex strings in the SUSY Abelian Higgs model.
We summarize this paper in section~\ref{sum}.
We use the notation and convention of the 
textbook~\cite{Wess:1992cp}.

\section{Dual transformations of vortex strings in Abelian Higgs model}
\label{AH}

In this section, we review two dual transformations of 
the ANO vortex strings of the bosonic Abelian Higgs model~\cite{Sugamoto:1978ft,Lee:1993ty}
at a classical level.
One is the transformation to the system described by a 1-form gauge 
field and a 2-form gauge field, where the ANO vortex strings are 
electrically coupled with the 2-form gauge field.
The other is the transformation to the system 
described by a massive 2-form field, which is also coupled with 
the ANO vortex strings.
\subsection{Abelian Higgs model}
Here, we introduce the Lagrangian of the Abelian Higgs model.
The Lagrangian is given by
\begin{equation}
 {\cal L}_{\rm AH} 
 = -\lb| \der_m \phi -\fr{e}{2}iA_m \phi\rb|^2 
-\fr{1}{4}F^{mn} F_{mn} 
- \fr{1}{8}(e|\phi|^2 - \xi)^2.
\label{181230.1758}
\end{equation}
Here, $A_m$  ($m = 0,1,2,3$) is a $U(1)$ gauge field, 
$F_{mn} = \der_m A_n -\der_n A_m$ is the field strength 
of the gauge field,
$\phi$ is a complex scalar field with the $U(1)$ charge $e/2$,
$e$ is a positive coupling constant of the $U(1)$ gauge field, 
$\xi $ is a positive parameter of mass-dimension two.
Note that the parameters are normalized 
so that the model can be embedded into SUSY theories.
The vacuum of the model is given by the 
minimum of the potential where $|\phi|$ develops non-zero 
vacuum expectation value:
\begin{equation}
 |\phi|^2 = \fr{\xi}{e}.
\end{equation}
Therefore, the $U(1)$ symmetry is spontaneously 
broken in this vacuum.
The vacuum is in the Higgs phase since 
the gauge field becomes massive by eating the phase 
of the scalar field.

\subsection{Dual 2-form gauge theory with vortex strings}
\label{bST}
The Higgs phase admits spatial dimension one (codimension two)
objects, since the first homotopy group of the vacuum manifold is 
nontrivial: $\pi_1 (U(1)) =\bb{Z}$.
The extended objects are so-called ANO vortex strings.
The positions of the ANO vortex strings are characterized by 
the zero points of $\phi$, where the $U(1)$ symmetry is 
recovered.

In the Lagrangian, 
the ANO vortex strings are expressed by using 
multivalued part of the phase of the complex scalar field (see 
e.g., Ref.~\cite{Kleinert:2008zzb}).
We split the complex scalar field as follows:
\begin{equation}
 \phi =\fr{1}{\sr{2}} \rho  e^{i(\varphi + \varphi_{\rm 0})}.
\end{equation}
Here, $\rho$ and $\varphi$ are real single-valued scalar fields,
 and $\varphi_0$ is 
a real multivalued scalar field.
In general, the phase can be multivalued since 
$\varphi_0 \to \varphi_0 + 2\pi $ 
does not change $\phi$.
The Lagrangian in \er{181230.1758} can be  
rewritten as
\begin{equation}
 \begin{split}
{\cal L}_{\rm AH}
 & = -\lb| \der_m \phi - i\fr{e}{2} A_m\phi\rb|^2 
-\fr{1}{4}F^{mn}F_{mn} 
+\cdots
\\
&=
-\fr{1}{2}\lb| \der_m \rho
+ i (\der_m \varphi + \der_m \varphi_0)\rho
- i\fr{e}{2} A_m\rho\rb|^2 
-\fr{1}{4}F^{mn}F_{mn} 
+\cdots\\
&=
-\fr{1}{2}(\der_m\rho)^2
-\fr{1}{2}\rho^2
\(\der_m\varphi +\der_m \varphi_0- \fr{e}{2}A_m \)^2
-\fr{1}{4}F^{mn}F_{mn} 
+\cdots,
 \end{split}
\label{181227.2110}
\end{equation}
where the ellipsis $\cdots$ refers to the terms which are
irrelevant to the dual formulations.

We dualize the scalar field $\varphi$ to a 2-form gauge field
as follows.
We introduce the following first-order Lagrangian
which is classically equivalent to the Lagrangian 
in \er{181227.2110}:
\begin{equation}
{\cal L}_{B,{\rm 1st}} = -\fr{1}{2}\rho^2
\(C_m +\der_m \varphi_0- \fr{e}{2} A_m \)^2
 + \fr{1}{2!}\epsilon^{mnpq} B_{mn}  \der_p C_q 
-\fr{1}{4}F^{mn}F_{mn} ,
\label{181227.2144}
\end{equation}
where we have omitted the terms which are irrelevant to 
the following discussions.
Here, $C_m $ is a 1-form gauge field without singularities, 
and $B_{mn}$ is a 2-form gauge field.
The gauge field $B_{mn}$ is transformed as 
$B_{mn} \to B_{mn}+ \der_m \lambda_n - \der_n \lambda_m$,
where $\lambda_m$ is a 1-form gauge parameter.
The gauge field $C_m$ is transformed under the 
gauge transformation of $A_m \to A_m + \der_m u$
as $C_m \to C_m + \fr{e}{2} \der_m u$, where $u$ is a 
gauge parameter.
The equivalence between the Lagrangian and the one 
in \er{181227.2110} can be seen 
by solving the EOM for $B_{mn}$, which gives us $C_m = \der_m \varphi$.
The dual formulation can be obtained by using the EOM for $C_m $ and by eliminating the field. 
The EOM for the $C_m$ gives us
\begin{equation}
C_m
=
  \fr{1}{\rho^2}(*H)_m 
-\der_m \varphi_0 + \fr{e}{2}A_m ,
\end{equation}
where we have defined
\begin{equation}
 H_{mnp} := \der_m B_{np} + \der_n B_{pm} + \der_p B_{mn}
\end{equation}
and
\begin{equation}
(*H)^m := \fr{1}{3!} \epsilon^{mnpq} H_{npq}.
\end{equation}
Therefore, the first-order Lagrangian in \er{181227.2144} 
becomes
\begin{equation}
\begin{split}
{\cal L}_{B} 
&= \fr{1}{2\rho^2} (*H)^m(*H)_m
-\fr{1}{4}F^{mn}F_{mn} 
+ \fr{e}{2}\cdot \fr{1}{2! \cdot 2! }\epsilon^{mnpq}B_{mn} F_{pq}
\\&\quad
 - \fr{1}{2!}\epsilon^{mnpq} B_{mn} \der_p \der_q \varphi_0.
\end{split}
\label{181227.2145} 
\end{equation} 
The Lagrangian describes a system with 
1-form and 2-form gauge field with the topological 
coupling $\epsilon^{mnpq} B_{mn} F_{pq}$.
The 2-form gauge field is electrically coupled with 
the string current $\epsilon^{mnpq} \der_p\der_q \varphi_0$.
Naively, $\epsilon^{mnpq} \der_p\der_q \varphi_0$ seems 
to be identically zero.
However, since $\varphi_0$ is a multivalued function,
$\epsilon^{mnpq} \der_p\der_q \varphi_0$ is non-zero
where $|\phi|$ becomes zero.
This can be seen as follows.
Since $\varphi_0$ is a part of the phase of the complex scalar field,
$\epsilon^{mnpq} \der_p\der_q \varphi_0$ can be rewritten as follows:
\begin{equation}
 \epsilon^{mnpq} \der_p\der_q \varphi_0 
=  \fr{1}{2 i} \epsilon^{mnpq} \der_p\der_q \log(\phi/\b\phi).
\label{181228.1536}
\end{equation}
Here, we have included the regular part $\varphi$ 
since $ \epsilon^{mnpq} \der_p\der_q \varphi =0$.
The right hand side of \er{181228.1536} gives rise to a delta function:
\begin{equation}
\begin{split}
\fr{1}{2 i} \epsilon^{mnpq} \der_p\der_q \log(\phi/\b\phi)
 &= 
\fr{1}{2 i} \epsilon^{mnpq} 
\der_p \(\fr{1}{\phi}\der_q \phi  - \fr{1}{\b\phi} \der_q \b\phi\) 
\\
 &= 
-2\pi i \epsilon^{mnpq} \delta^2(\phi,\b\phi)
\der_p \b\phi\der_q \phi. 
\end{split}
\label{181228.1546}
\end{equation}
Here, we have used a property of a two dimensional delta function
\begin{equation}
\pd{}{\b\phi} \fr{1}{\phi} 
= 
\pd{}{\phi} \fr{1}{\b\phi} 
= 2\pi \delta^2 (\phi, \b\phi),
\label{181231.1001}
\end{equation}
where the delta function is defined by 
\begin{equation}
 \delta^2(\phi, \b\phi) 
:= \fr{1}{2} \delta({\rm Re}\,\phi)  \delta({\rm Im}\,\phi).
\label{181231.1002}
\end{equation}
Thus, $\epsilon^{mnpq}\der_p \der_q\varphi_0$ has 
singularities of the delta function where $|\phi|$ is zero.
The string current is a conserved current:
\begin{equation}
 \der_n \epsilon^{mnpq} \der_p \der_q \varphi_0
 = 0,
\end{equation}
because 
$\epsilon^{mnpq} \der_n \phi \der_p \phi \der_q\b\phi = 
\epsilon^{mnpq} \der_n \phi \der_p \b\phi \der_q\b\phi =0 $.
\subsection{Dual massive 2-form theory with vortex strings}

We have dualized the Lagrangian in \er{181227.2110}
into the one with the 2-form gauge field.
We can also dualize the Lagrangian 
into a system with a massive 2-form field.
We introduce the following first-order Lagrangian
which is classically equivalent to the Lagrangian in \er{181227.2145}:
\begin{equation}
\begin{split}
{\cal L}_{B',{\rm 1st}} 
&= 
\fr{1}{2\rho^2} (*H)^m(*H)_m
 -\fr{1}{2!}\epsilon^{mnpq} B_{mn} \der_p\der_q \varphi_0
+ \fr{e}{2}\cdot\fr{1}{3!}\epsilon^{mnpq} A_m H_{npq}
\\
&
\quad
-\fr{1}{4}F'^{mn}F'_{mn} 
+ \fr{1}{2!\cdot 2!}
\epsilon^{mnpq} B'_{mn} (\der_p A_q -\der_q A_p -F'_{pq}). 
\end{split}
\label{181227.2146} 
\end{equation} 
Here, $B'_{mn}$ is a 2-form field as a Lagrange's multiplier,
$F'_{mn}$ is a 2-form field which is independent of the 
original 1-form gauge field $A_m$.
The EOM for the Lagrange's multiplier gives us the 
relation $F'_{mn} = \der_m A_n - \der_n A_m$, and 
we go back to the original Lagrangian in \er{181227.2145}.
Instead,
the EOM for $A_m$ gives us 
\begin{equation}
  \fr{1}{3!} \epsilon^{mnpq}H_{npq}
 = -\fr{2}{e} \cdot \fr{1}{2!}\epsilon^{mnpq} \der_n B'_{pq}.
\label{181229.2046} 
\end{equation}
Furthermore, the EOM for $F'_{mn}$ leads to 
\begin{equation}
  F'_{mn} = - \fr{1}{2!}\epsilon_{mnpq} B'^{pq}. 
\label{181229.2047}
\end{equation}
Substituting \ers{181229.2046} and \eqref{181229.2046} into 
the Lagrangian in \er{181227.2146}, 
we obtain
\begin{equation}
\begin{split}
 {\cal L}_{B'} 
&= 
\fr{2}{e^2 \rho^2} (*H')^m(*H')_m
-\fr{1}{4}B'^{mn}B'_{mn} 
 -\fr{2}{e}\cdot 
\fr{1}{2!}\epsilon^{mnpq} B'_{mn} \der_p\der_q \varphi_0
\end{split}
\label{181229.2051}
\end{equation}
up to total derivatives.
Here, we have defined 
\begin{equation}
 (*H')^m = \fr{1}{2!} \epsilon^{mnpq} \der_n B'_{pq}.
\end{equation}
The second term of the Lagrangian in \er{181229.2051} is 
the mass term for the 2-from $B'_{mn}$.
Therefore, the Lagrangian in \er{181229.2051} describes a system of 
massive 2-form field.
The ANO vortex strings are coupled with the massive 2-form.

\section{Dual transformations of vortex strings in SUSY Abelian Higgs model}
\label{SAH}

In this section, we discuss the dual transformations 
of ANO vortex strings of the SUSY Abelian Higgs model.
In SUSY theories, the Higgs potential can be obtained by
a F-term or a D-term potentials~\cite{Davis:1997bs}.
For the former case, 
the SUSY is completely broken in 
the core of the vortex strings.
For the latter case, 
the half of SUSY can be preserved 
in the core of the ANO vortex strings,
 and the ANO vortex strings can be BPS states~\cite{Davis:1997bs,Dvali:2003zh,Copeland:2003bj}.
We thus discuss the latter option in this paper.

We use the superspace formalism in order to obtain 
the manifestly SUSY theories.
The superspace is spanned by the coordinates 
$(x^m,\theta^\alpha,\b\theta_{\d\alpha})$,
 where $(x^m) $ ($m = 0,1,2,3$) are coordinates of 
the Minkowski spacetime, and $(\theta^\alpha,\b\theta_{\d\alpha})$
are coordinates spanned by the Grassmann numbers.
The indices beginning with $m,n, ... $
are vector indices.
The indices beginning with $\alpha, \beta,... \d\alpha,\d\beta,...$
are spinor indices with $\alpha = 1,2$ and $\d\alpha = \d{1},\d{2}$.

\subsection{SUSY Abelian Higgs model}
We introduce a Lagrangian of the SUSY Abelian Higgs model.
We begin with the following Lagrangian:
\begin{equation}
{\cal L}_{\rm AH,SUSY}
=
\fr{1}{2}\int d^4\theta (\b\Phi e^{eV} \Phi 
+ \b{\tilde\Phi} e^{-eV} \tilde\Phi
- \xi V) 
+ \fr{1}{4}\int d^2\theta W^\alpha W_\alpha 
+\hc
\label{181225.1759} 
\end{equation}
Here, 
$V$ is a vector superfield in which a $U(1)$ 
vector gauge field $A_m $ is embedded,
$W_\alpha = -\fr{1}{4} \b{D}^2 D_\alpha V$ is a 
gaugino superfield given by the vector superfield, 
 $e$ is a positive coupling constant of the $U(1)$ gauge symmetry, 
and $\xi$ is a Fayet--Iliopoulos (FI) parameter~\cite{Fayet:1974jb}.
Superfields $\Phi$ and $\tilde\Phi$ are chiral superfields
with $U(1)$ charge $+e/2$ and $-e/2$, respectively.
The chiral superfields are transformed by 
the $U(1)$ gauge transformation as 
$\Phi \to \Phi e^{e\Lambda}$ and 
$\tilde\Phi \to \tilde\Phi e^{-e\Lambda}$
when 
$V$ is transformed as 
$V \to V - \Lambda -\b\Lambda$.
Here, $\Lambda$ is a chiral superfield parameter.
The bosonic part of the component Lagrangian is 
\begin{equation}
\begin{split}
{\cal L}_{\rm AH,SUSY,boson}
&=
- \lb|\der_m \phi -i\frac{e}{2} A_m \phi\rb|^2  
- \lb|\der_m \tilde\phi +i\frac{e}{2} A_m \tilde\phi\rb|^2  
-\fr{1}{4}F^{mn}F_{mn} 
\\
&\quad
+\fr{1}{2} \bs{D} (e|\phi|^2 - e|\tilde\phi|^2 - \xi)
+F\b{F} + \tilde{F}\b{\tilde{F}}
+ \fr{1}{2} \bs{D}^2.
\end{split}
\label{181213.1823}
\end{equation}
Here, we have omitted fermions which are not needed for the 
following discussion in this section.
In the Lagrangian \er{181213.1823}, we have used the Wess--Zumino (WZ)
gauge: 
$V| = D_\alpha V |= \b{D}_{\d\alpha } V |= D^2 V| = \b{D}^2 V | =0$. 
Here, the vertical bar ``$|$'' represents $\theta = \b\theta =0$
projection of the superfields, and 
$D_\alpha$ and $\b{D}_{\d\alpha} $ are SUSY covariant spinor 
derivatives.
The components of the chiral superfield $\Phi$ and 
the vector superfield $V$ are denoted as
\begin{equation}
\phi = \Phi|, 
\quad
\chi_\alpha = \fr{1}{\sr{2}} D_\alpha \Phi|,
\quad
F = -\fr{1}{4} D^2 \Phi|, 
\end{equation}
\begin{equation}
\tilde\phi = \tilde\Phi|, 
\quad
\tilde\chi_\alpha = \fr{1}{\sr{2}} D_\alpha \tilde\Phi|,
\quad
\tilde{F} = -\fr{1}{4} D^2 \tilde\Phi|, 
\end{equation}
\begin{equation}
\begin{split}
&A_{\alpha\d\alpha} =  \fr{1}{2}[D_\alpha,\b{D}_{\d\alpha}] V|,
\\
&
F_{mn} = \der_m A_n  -\der_n A_m = 
\fr{1}{2i} 
(
(\sigma_{mn})_\alpha{}^\beta D^\alpha W_\beta
-
(\b\sigma_{mn})^{\d\alpha}{}_{\d\beta} \b{D}_{\d\alpha} \b{W}^{\d\beta}) |,
\\
&
\lambda_\alpha = iW_\alpha |, 
\quad 
\b\lambda^{\d\alpha} = -i\b{W}^{\d\alpha}| ,
\quad
\bs{D} = -\fr{1}{2}D^\alpha W_\alpha|
 = -\fr{1}{2} \b{D}_{\d\alpha} \b{W}^{\d\alpha}|.
\end{split}
\end{equation}
The quantities
$(\sigma^m)_{\alpha\d\alpha}$ and 
$(\b\sigma^m)^{\d\alpha\alpha} $
are four-dimensional Pauli 
matrices which satisfy
$ (\b\sigma^m)^{\d\alpha\alpha} = (\sigma^m)^{\alpha\d\alpha}$.
The quantity $A_{\alpha\d\alpha}$ 
is defined by the Pauli matrices as 
$A_{\alpha\d\alpha} = (\sigma^m)_{\alpha\d\alpha} A_m$.
The quantities 
$(\sigma^{mn})_\alpha{}^\beta $ and $(\b\sigma^{mn})^{\d\alpha}{}_{\d\beta} $ are self-dual and anti-self dual tensors defined by
\begin{equation}
 \begin{split}
 (\sigma^{mn})_\alpha{}^\beta 
&= \fr{1}{4}((\sigma^m)_{\alpha\d\gamma} (\b\sigma^n)^{\d\gamma\beta}
 - (\sigma^n)_{\alpha\d\gamma} (\b\sigma^m)^{\d\gamma\beta}),
\\
(\b\sigma^{mn})^{\d\alpha}{}_{\d\beta} 
&= \fr{1}{4}((\b\sigma^m)^{\d\alpha\gamma} (\sigma^n)_{\gamma\d\beta}
 - (\b\sigma^n)^{\d\alpha\gamma} (\sigma^m)_{\gamma\d\beta}),
 \end{split}
\end{equation}
respectively.

In this model, the $U(1)$ symmetry is spontaneously broken
if the FI parameter $\xi$ is non-zero.
This can be seen by the on-shell potential ${\cal V}$
\begin{equation}
 {\cal V} =
-\fr{1}{2} \bs{D} (e|\phi|^2 - e|\tilde\phi|^2 - \xi)
-F\b{F} - \tilde{F}\b{\tilde{F}}
- \fr{1}{2} \bs{D}^2 .
\end{equation}
In order to obtain the on-shell potential, we solve 
the EOM for the auxiliary fields $F$ and $\bs{D}$.
The EOM for $F$ and $\tilde{F}$ are trivial: $F = \tilde{F} = 0$,
while the EOM for $\bs{D}$ is 
\begin{equation}
 \bs{D} =  - \fr{1}{2}(e|\phi|^2 -e |\tilde\phi|^2 - \xi).
\end{equation}
Therefore, the on-shell potential ${\cal V}$ is 
\begin{equation}
 {\cal V} = \fr{1}{8}(e|\phi|^2 -e |\tilde\phi|^2 - \xi)^2.
\end{equation}
The vacuum of the model is given by the 
minimum of the potential, 
which is described by the condition
\begin{equation}
 e|\phi|^2 -e |\tilde\phi|^2  = \xi.
\end{equation}
If the FI parameter is positive $\xi >0$, $|\phi|^2$ cannot 
be zero while $|\tilde\phi|^2$ can be zero.
Since $\phi$ develops the vacuum expectation value, 
the $U(1)$ symmetry is broken,
 and the vector field $A_m$ becomes massive by 
eating the phase of $\phi$.
Note that SUSY is unbroken in this vacuum since  
the vacuum expectation value of
the auxiliary field is $ \bs{D} = 0$ in this vacuum.

\subsection{Dual SUSY 2-form gauge theory with vortex strings}
\label{ST}
We consider a dual formulation of the 
SUSY Abelian Higgs model.
We use the superspace formalism in order to make SUSY manifest.
In section~\ref{bST}, we have reviewed the 
dual transformations of the bosonic Abelian Higgs model.
As in the bosonic Abelian Higgs model,
there are at least two ways to dualize the Lagrangian.
One is to dualize the chiral superfield $\Phi$.
In this case, the dual theory is described by 
a 2-form gauge field $B_{mn}$ in addition 
to the original 1-form gauge field $A_m $.
In the dual theory, the 2-form gauge field is 
topologically coupled with
the 1-form gauge field.
The other is to dualize the vector superfield $V$.
In this case, the dual theory is described by 
a massive 2-form field where the 1-form gauge field is eaten by 
the 2-form gauge field.
In this subsection, we choose the former option.
The ANO vortex strings are coupled with the 2-form gauge field 
electrically. 
\subsubsection{String current superfield} 
We begin with the following Lagrangian:
\begin{equation}
{\cal L}'_{\rm AH,SUSY}
=
\fr{1}{2}\int d^4\theta (\b\Phi e^{eV} \Phi - \xi V) 
+ \fr{1}{4}\int d^2\theta W^\alpha W_\alpha 
+\hc,
\label{181001.2225}
\end{equation}
where we have omitted the
terms which are irrelevant to
the ANO vortex strings,
since we are interested in the dual formulation of the ANO vortex 
strings.
In the presence of the vortex strings, 
the Lagrangian has singular points in 
the field space of $\Phi$ where 
$\Phi = 0$.
In order to dualize the Lagrangian,
we split $\Phi$
into the singular part and the regular 
part as follows:
\begin{equation}
\Phi= \Phi_0 \Phi_1,
\qtq{where} \Phi_0 := \fr{\Phi}{\Phi_1}.
\end{equation}
Here, $\Phi_1$ is a regular chiral superfield of 
mass-dimension one,
which does not have a zero-point.
This regular part can be understood as a SUSY extension of 
$e^{i\varphi}$ in the bosonic model in section~\ref{bST}.
Since $\Phi_1$ is the regular chiral superfield, 
we can assign non-singular gauge transformations for $\Phi_1$.
We assume the same gauge transformation law of the chiral superfield 
$\Phi_1$ as $\Phi$: 
$\Phi_1 \to \Phi_1 e^{e\Lambda}$.
Since $\Phi_1$ is not zero everywhere, there are no singular points for the gauge transformation. 
On the other hand, 
$\Phi_0$ has singular points where 
$\Phi_0$ is zero.
Again, this singular part can be understood as a SUSY extension of 
$e^{i\varphi_0}$ in the bosonic model in section~\ref{bST}.
This zero-point is originated from 
the zero-point of the chiral superfield
$\Phi$.
Thanks to the splitting $\Phi=  \Phi_0 \Phi_1$,
we can discuss the regular and singular parts in a manifestly gauge covariant and invariant ways, respectively. 

We rewrite the Lagrangian in \er{181213.1823} 
by using $\Phi_0$ and $\Phi_1$ as follows:
\begin{equation}
{\cal L}'_{\rm AH,SUSY}
 = \fr{1}{2}\int d^4\theta |\Phi_0|^2 
 |\Phi_1|^2 e^{eV} 
 + \fr{1}{4}\int d^2\theta W^\alpha W_\alpha
-\fr{1}{2} \xi \int d^4\theta V + \hc
\end{equation}
Now, we dualize $|\Phi_1|^2$ by the following
the first-order Lagrangian:
\begin{equation}
\begin{split}
{\cal L}'_{B,{\rm SUSY,1st}}
&= \fr{1}{2}\int d^4\theta |\Phi_0|^2 M^2 
e^{U+eV}
+\fr{1}{4}\int d^2\theta W^\alpha W_\alpha
-\fr{1}{2} \xi \int d^4\theta V
\\
&\quad
- \fr{1}{4\cdot 2i} \int d^2 \theta 
\Sigma^\alpha \b{D}^2 D_\alpha U 
+ \hc
\end{split}
\label{181223.1806}
\end{equation}
Here, $U$ is a real superfield whose 
gauge transformation law is 
$U \to U +e (\Lambda+ \b\Lambda)$ under $V \to V -\Lambda  -\b\Lambda$.
The superfield $\Sigma_\alpha$ is a chiral superfield,
 $M$ is a parameter of mass-dimension one.
Since the original chiral superfield $\Phi_1$ is regular, we can safely assume 
that $U$ is also a regular function
in the sense that 
$e^U$ does not have zero points.
The Lagrangian is invariant under the gauge transformation 
of $\Sigma_\alpha$:
\begin{equation}
 \delta_2 \Sigma_\alpha  = -\fr{1}{4} \b{D}^2 D_\alpha \Theta,
\label{181231.1020}
\end{equation}
where $\delta_2$ refers to an infinitesimal gauge transformation 
of $\Sigma_\alpha$, and $\Theta$ is a real superfield parameter.
Since the chiral spinor superfield with the 
gauge transformation in \er{181231.1020} includes the 
2-form gauge field $B_{mn}$ as a component field (see e.g.,~Ref.~\cite{Gates:1980ay}), 
we call $\Sigma_\alpha$ ``2-form prepotential''
following Ref.~\cite{Gates:1983nr}.

We can go back to the original Lagrangian in \er{181213.1823}
by eliminating $\Sigma_\alpha$ by its EOM.
The EOM for $\Sigma_\alpha$ and 
its Hermitian conjugate,
\begin{equation}
\b{D}^2 D_\alpha U = D^2 \b{D}_{\d\alpha} U=0,
\end{equation}
give us the following solution:
 \begin{equation}
 U = \Phi' +\b\Phi',
 \end{equation}
where $\Phi'$ is a single valued chiral 
superfield since $e^U$ is a non-zero superfield.
If we define $\Phi_1 = e^{\Phi'}$,
we obtain the original Lagrangian.

The dual formulation can be obtained by eliminating 
the real superfield $U$ instead of eliminating $\Sigma_\alpha$.
The EOM for $U$ is 
\begin{equation}
0= |\Phi_0|^2 M^2 e^{U + eV} - L,
\end{equation}
where $L$ is a real superfield 
defined by
\begin{equation}
    L = \fr{1}{2i} 
(D^\alpha \Sigma_\alpha
    - \b{D}_{\d\alpha} \b\Sigma^{\d\alpha}).
\end{equation}
Note that the real superfield $L$ is a linear superfield 
since $D^2 L  =\b{D}^2 L = 0$.

By using the real linear superfield $L$, 
$U$ can be solved as
\begin{equation}
U = \log \fr{L}{|\Phi_0|^2 M^2 e^{eV}}.
\end{equation}
Substituting the solution into 
the first-order Lagrangian in \er{181223.1806}, 
we reach at the following dual Lagrangian 
\begin{equation}
\begin{split}
{\cal L}'_{B,{\rm SUSY}}
&= 
-\fr{1}{2} \int d^4 \theta 
L\log \(\fr{L}{M^2}\)
+\fr{1}{4}\int d^2\theta W^\alpha W_\alpha
-\fr{1}{2}\int d^4\theta  \xi  V
\\
&
\quad
-\fr{e}{2i}\int d^2\theta \Sigma^\alpha W_\alpha
-\fr{1}{2i}\int d^2\theta \Sigma^\alpha J_\alpha+ \hc    
\end{split}
\label{181201.2036}
\end{equation}
Here, we have defined the chiral superfield $J_\alpha$ as
\begin{equation}
J_\alpha := -\fr{1}{4} \b{D}^2 D_\alpha 
\log |\Phi_0|^2,
\end{equation}
and we call the superfield $J_\alpha$ ``string current superfield''
for the later convenience.
The terms $\fr{1}{2i}\int d^2\theta \Sigma^\alpha J_\alpha+ \hc$ 
are invariant under the gauge transformation of the 2-form prepotential
in \er{181231.1020}, because
the chiral superfield $J_\alpha$ satisfies 
the following identity like the gaugino superfield
\begin{equation}
D^\alpha J_\alpha = \b{D}_{\d\alpha}\b{J}^{\d\alpha}
\label{181229.2209}
\end{equation}
by the SUSY algebra 
$D^\alpha \b{D}^2 D_\alpha  = \b{D}_{\d\alpha} D^2 \b{D}^{\d\alpha}$.
Naively, $J_\alpha =0 $ since 
$\log |\Phi_0|^2 = \log \Phi_0 + \log \b\Phi_0$.
However, since $\Phi_0$ can have zero points, 
$\b{D}^2 D_\alpha \log |\Phi_0|^2$
contains a singularity of a delta function.
We will discuss the singularity more precisely.
\subsubsection{Component expression of dual formulation}

In the Lagrangian in \er{181201.2036}, 
there are a coupling between the 2-form 
gauge field and the string current and its SUSY completion.
The coupling and its SUSY completion are given by the last term.
To see the coupling, we express 
the dual Lagrangian ${\cal L}'_{B,{\rm SUSY}}$ 
in terms of the component fields.
The component expression is 
\begin{equation}
\begin{split}
{\cal L}'_{B,{\rm SUSY}}
&=
-\fr{1}{2\sr{2}\sigma}
\(
(\der^m \sigma) (\der_m \sigma) 
-(*H)^m (*H)_m\)
\\
&\quad
-\fr{i}{2\sr{2}\sigma}
\(
\b{\psi}_{\d\alpha} 
(\b\sigma^m)^{\d\alpha\alpha} \der_m \psi_\alpha 
+
\psi^\alpha  (\sigma^m)_{\alpha\d\alpha}\der_m \b{\psi}^{\d\alpha}
\)
\\
&
\quad
- \fr{1}{4\sigma^2}
\psi^\alpha (\b\sigma^m)_{\alpha\d\alpha} \b{\psi}^{\d\alpha} 
 (*H)_m
-
\fr{1}{4\sr{2}\sigma^3} 
\psi^\alpha\psi_\alpha \b\psi_{\d\alpha}\b\psi^{\d\alpha}
\\
&
\quad
-\fr{1}{4}F^{mn}F_{mn}
-\fr{i}{2}(\b\lambda^\alpha (\sigma^m)_{\alpha\d\alpha} \der_m \b\lambda^{\d\alpha}
+ \b\lambda_{\d\alpha} (\b\sigma^m)^{\d\alpha\alpha}\der_m \lambda_\alpha) 
+ \fr{1}{2} \bs{D}^2
-\fr{1}{2} \xi \bs{D}
\\
&\quad
+
\fr{e}{2\cdot 2!\cdot 2!}  \epsilon^{mnpq} B_{mn} F_{pq}  
+\fr{ie}{\sr{2}}
(\lambda^\alpha \psi_\alpha
-\b\lambda_{\d\alpha}\b{\psi}^{\d\alpha })
+2 \sr{2} e\sigma \bs{D}
\\
&
\quad
-
\fr{1}{\sr{2}} 
(\psi^\alpha j_\alpha +\b\psi_{\d\alpha}\b{j}^{\d\alpha})
+\fr{1}{2\cdot 2!} \tilde{J}^{mn} B_{mn}
+\sr{2} \sigma J.
\end{split}
\label{181214.1609}
\end{equation}
Here, 
the components of the chiral superfield $\Sigma_\alpha$ and 
the linear sueprfield $L$
are denoted as
\begin{equation}
\begin{split}
&
B_{mn} = 
-i
((\sigma_{mn})_\alpha{}^\beta D^\alpha \Sigma_\beta
 - (\b\sigma_{mn})^{\d\alpha}{}_{\d\beta} \b{D}_{\d\alpha} \b\Sigma^{\d\beta}),
\\
&H_{mnp} =
\der_m B_{np} + \der_n B_{pm} + \der_p B_{mn}
= \fr{1}{4} \epsilon_{mnpq} 
(\b\sigma^q)^{\d\alpha\beta}[D_\beta,\b{D}_{\d\alpha}] L,
\\
& \sigma : = \fr{1}{\sr{2}}L|, 
\quad
\psi_\alpha :=  \fr{1}{\sr{2}} D_\alpha L| 
= +\fr{i}{4\sr{2}}D^2\Sigma_\alpha|, 
\quad
\b\psi_{\d\alpha} :=  
\fr{1}{\sr{2}} \b{D}_{\d\alpha}L |
= -\fr{i}{4\sr{2}}\b{D}^2\b\Sigma_{\d\alpha}|.  
\end{split}
\end{equation}
The vector component can also be written as
\begin{equation}
(*H)^m =
\fr{1}{3!}\epsilon^{mnpq}H_{npq}  
= \fr{1}{4} 
(\b\sigma^m)^{\d\alpha\beta}[D_\beta,\b{D}_{\d\alpha}] L,
 \end{equation}
or
\begin{equation}
[D_\alpha,\b{D}_{\d\alpha}] L|
 = 
-2
(*H)_{\alpha\d\alpha} .
\end{equation}
Note that we have used the WZ gauge for the 2-form prepotential
$\Sigma_\alpha$:
\begin{equation}
 \Sigma_\alpha| = \b\Sigma_{\d\alpha}| 
= (D^\alpha \Sigma_\alpha + \b{D}_{\d\alpha} \b\Sigma^{\d\alpha})| =0.
\end{equation}
Note that the superparters of the phase of the complex scalar 
field are also dualized to the 2-form prepotential 
in the Lagrangian in \er{181214.1609} due to SUSY
in contrast to the bosonic case.
In the Lagrangian in \er{181214.1609},
the fields $j_\alpha$, $\b{j}_{\d\alpha}$, $J_{mn}$, and $J$ 
are the components of $J_\alpha$:
\begin{equation}
\begin{split}
 &
 j_\alpha = J_\alpha |, 
\quad
\b{j}_{\d\alpha} = \b{J}_{\d\alpha}|,
\\
&J_{mn} = \fr{1}{2i}((\sigma_{mn})_\alpha{}^\beta D^\alpha J_\beta
 - 
(\b\sigma_{mn})^{\d\alpha}{}_{\d\beta} 
\b{D}_{\d\alpha} \b{J}^{\d\beta})|,
\quad
\tilde{J}_{mn} = \fr{1}{2!}\epsilon_{mnpq}J^{pq},
\\
&
J = -\fr{1}{2} D^\alpha J_\alpha| = -\fr{1}{2} \b{D}_{\d\alpha}\b{J}^{\d\alpha}|.
\end{split}
\end{equation}
It seems that $J_\alpha =0 $ since 
$\log |\Phi_0|^2 = \log \Phi_0 + \log \b\Phi_0$.
However, since $\Phi_0$ can have zero points,
 $\b{D}^2 D_\alpha \log |\Phi_0|^2$
should contain a term like a delta function as mentioned above.
The delta function arises as 
a SUSY extension of \er{181231.1001}:
\begin{equation}
 \pd{}{\Phi_0} \pd{}{\b\Phi_0} \log |\Phi_0|^2 
= 
 \pd{}{\Phi_0} \fr{1}{\b\Phi_0}
=  
 \pd{}{\b\Phi_0} \fr{1}{\Phi_0}
= 2\pi \delta^2 (\Phi, \b\Phi),
\end{equation}
where $\delta^2(\Phi,\b\Phi)$ is defined by
\begin{equation}
\delta^2 (\Phi,\b\Phi) = 
\fr{1}{2}\delta({\rm Re}\,\Phi)\delta({\rm Im}\,\Phi) .
\end{equation}
Note that this property of $\log |\Phi_0|^2$ 
can also be understood as
a SUSY extension of the 
two-dimensional Green's function.
We explicitly write down the components of $J_\alpha$ as follows:
\begin{equation}
 \begin{split}
j_\alpha &= J_\alpha |
 = -\fr{1}{4}
\b{D}^2 D_\alpha \log |\Phi_0|^2|
\\
&= 
2\sr{2}
\pi \delta^2 (\phi_0,\b\phi_0)
 \b{F}_0 \chi_{0\alpha} 
-2\sr{2}i
\pi \delta^2 (\phi_0,\b\phi_0)
  (\b\sigma^m)_{\alpha \d\alpha} \der_m\phi_0 
\b{\chi}^{0\d\alpha}
\\
&
\quad
-\sr{2}
\pi \(\pd{}{\b\phi_0} \delta^2 (\phi_0,\b\phi_0)\)
 \b{\chi}_{0\d\alpha}\b{\chi}^{\d\alpha}_0 \chi_{0\alpha} .
 \end{split}
\end{equation}
Here, $\phi_0 $, $\chi_0$, and $F_0$ are defined by
\begin{equation}
 \phi_0 = \Phi_0|, \quad
\chi_{0\alpha} = \fr{1}{\sr{2}} D_\alpha \Phi_0 |,
\quad
F_0 = -\fr{1}{4} D^2 \Phi_0|,
\end{equation}
and we have used 
\begin{equation}
 D_\alpha \log |\Phi_0|^2
 =  (D_\alpha \Phi_0)\pd{}{\Phi_0} \log |\Phi_0|^2
\end{equation}
and its Hermitian conjugate.
The component $J$ can also be rewritten as
\begin{equation}
\begin{split}
&
J = -\fr{1}{2} D^\alpha J_\alpha |
= 
-\fr{1}{2}\b{D}_{\d\alpha} \b{J}^{\d\alpha}|
=
\fr{1}{8}
D^\alpha \b{D}^2 D_\alpha \log |\Phi_0|^2|
\\
&
=  
4
\pi \delta^2 (\phi_0,\b\phi_0)
\(-
  \der_m \phi_0 \der^m \b\phi_0
-\fr{i}{2}
\chi_0^\alpha 
(\sigma^n)_{\alpha\d\alpha} \der_n \b{\chi}_0^{\d\alpha}
-\fr{i}{2}
\b{\chi}_{0\d\alpha}  (\sigma^n)^{ \d\alpha\alpha} \der_n\chi_{0\alpha}
+
F_0 \b{F}_0
\)
\\
&
\quad
+2i
\pi \(
\pd{}{\b\phi_0} \delta^2 (\phi_0,\b\phi_0)
\der_m\b\phi_0 
-\pd{}{\phi_0}\delta^2 (\phi_0,\b\phi_0)
\der_m\phi_0
\)
\b{\chi}_{0\d\alpha}  (\b\sigma^m)^{\alpha\d\alpha}
 \chi_{0\alpha} 
\\
&
\quad
-2
\pi \(\pd{}{\b\phi_0} \delta^2 (\phi_0,\b\phi_0)\)
 \b{\chi}_{0\d\alpha} \b{\chi}^{\d\alpha}_0 F_0 
-2
\pi \(\pd{}{\phi_0}\delta^2 (\phi_0,\b\phi_0)\)
\chi_0^\alpha  \chi_{0\alpha}  \b{F}_0 
\\
&
\quad
+
\pi \(\pd{}{\phi_0}\pd{}{\b\phi_0} \delta^2 (\phi_0,\b\phi_0)\)
\chi^\alpha_0 \chi_{0\alpha} \b\chi_{0\d\alpha}\b\chi^{\d\alpha}_0
.
\end{split}
\end{equation}
This component may correspond to (the twice of) 
the Lagrangian of the non-linear sigma model where 
the K\"ahler potential is given by $K = \log |\Phi_0|^2$.
Finally, the component 
$\tilde{J}_{mn} = \fr{1}{2!} \epsilon_{mnpq} J^{pq}$ 
can be calculated as 
\begin{equation}
\begin{split}
\tilde{J}_{mn}
&=  
-
4i
\pi \delta^2 (\phi_0,\b\phi_0)
 \epsilon_{mnpq}
 \der^p \phi_0 \der^q \b\phi_0
\\
&
\quad
-
2\pi 
\delta^2 (\phi_0,\b\phi_0)
 \epsilon_{mnpq}
(\chi^\alpha_0 
(\sigma^q)_{\alpha\d\alpha}
(\der^p \b{\chi}^{\d\alpha}_0) 
-
\b{\chi}_{0\d\alpha}
(\b\sigma^q)^{\d\alpha\beta}
  \der^p \chi_{0\beta} 
)
\\
&
\quad
+
2\pi 
 \epsilon_{mnpq}
\(\pd{}{\b\phi_0} \delta^2 (\phi_0,\b\phi_0)
\der^p
\b\phi_0
+
\pd{}{\phi_0}\delta^2 (\phi_0,\b\phi_0)
 \der^p \phi_0
\)
 \b{\chi}_{0\d\alpha}  
(\b\sigma^q)^{\d\alpha\alpha}
\chi_{0\alpha} 
.
\end{split}
\label{181231.1203}
\end{equation}
Since the right hand side of the first line in \er{181231.1203}
corresponds to \er{181228.1546} in the bosonic case, 
$\tilde{J}_{mn}$ can be understood as
a SUSY extension of the string current.
The conservation law of $\tilde{J}_{mn}$ can be derived by
the relation $D^\alpha J_\alpha = \b{D}_{\d\alpha} \b{J}^{\d\alpha}$
in \er{181229.2209},
which implies
\begin{equation}
 \der_m \tilde{J}^{mn} =0.
\end{equation}
Note that \er{181229.2207} is equivalent to the property 
that $J_{mn}$ is closed:
\begin{equation}
 \epsilon^{mnpq} \der_n J_{pq} =0.
\label{181229.2207}
\end{equation}

Before closing this section,
a comment is in order on the string current superfield.
Since \ers{181229.2209} and \eqref{181229.2207} hold,  
the string current superfield can be a SUSY extension of the 
closed 2-form which cannot be expressed by the 
exterior derivative of a regular 
1-form.
If we regard 
the string current superfield as a ``gaugino superfield'' of 
a singular 2-form field strength, 
the ``prepotential'' for the gaugino superfield 
may correspond to $\log |\Phi_0|^2$.
In this case, the vector component of $\log |\Phi_0|^2$
is singular at the zero points of $\phi_0$:
\begin{equation}
([D_\alpha,\b{D}_{\d\alpha}] \log|\Phi_0|^2)  |
=
-2i (\sigma^m)_{\alpha\d\alpha}
\( \fr{1}{\b\phi_0}\der_m \b\phi_0
-\fr{1}{\phi_0}\der_m \phi_0 \)
-8\pi \delta^2 (\phi_0,\b\phi_0)
 \chi_{0\alpha}  \b\chi_{0\d\alpha}. 
\end{equation}
\subsection{Dual SUSY
massive 2-form theory with vortex strings}
\label{MT}
Here, we further dualize the Lagrangian in \er{181201.2036}.
The Lagrangian in \er{181201.2036} is described by 
the 1-form prepotential $V$ and the 2-form prepotential 
$\Sigma_\alpha$ with the topological coupling 
$\epsilon^{mnpq}B_{mn} F_{pq}$.
In this picture, the string current superfield is 
coupled with the 2-from prepotential in a gauge invariant way.
We can further dualize the Lagrangian as we will see below.
In this picture, the 2-form gauge field is manifestly massive 
by eating the 1-form gauge field.
The dual transformation 
can be done by adding a Lagrange's multiplier $\Upsilon_\alpha $ 
in the Lagrangian:
\begin{equation}
\begin{split}
{\cal L}'_{B',{\rm SUSY,1st}}
&= 
-\fr{1}{2} \int d^4 \theta 
L\log \(\fr{L}{M^2}\)
+\fr{1}{4}\int d^2\theta W'^\alpha W'_\alpha
+\fr{1}{2}\int d^4\theta (e L - \xi ) V
\\
&
\quad
-\fr{1}{2i}\int d^2\theta \Sigma^\alpha  J_\alpha
-\fr{1}{2i}
\int d^2 \theta \Upsilon^\alpha 
\(W'_\alpha  + \fr{1}{4} \b{D}^2 D_\alpha V\)
+ \hc
\end{split}
\label{181201.2037} 
\end{equation}
Here, $\Upsilon_\alpha$ is a chiral superfield as 
a Lagrange's multiplier,
 $W'_\alpha$ is a chiral superfield which is independent of 
the real superfield $V$.
Note that $\Upsilon_\alpha$ do not have a gauge symmetry 
in contrast to $\Sigma_\alpha$.
The EOM for $\Upsilon_\alpha $ gives us the original Lagrangian 
as before, while the EOM for $W'_\alpha$ gives us 
\begin{equation}
 W'_\alpha = -i \Upsilon_\alpha.
\label{181221.1415}
\end{equation}
This equation implies that $W'_\alpha $ is now described by 
the chiral superfield $\Upsilon'_\alpha$.
Further, the EOM for the 1-form prepotential $V$ leads to
\begin{equation}
 L =  \fr{1}{e}(\Psi+\xi),
\label{181221.1413}
\end{equation}
where $\Psi$ is given by
\begin{equation}
 \Psi := \fr{1}{2i} 
(D^\alpha \Upsilon_\alpha - \b{D}_{\d\alpha} \b\Upsilon^{\d\alpha}).
\end{equation}
The relation in \er{181221.1413} means that the 2-form prepotential 
$\Sigma_\alpha$ can be described by the chiral superfield 
$\Upsilon_\alpha$.
Substituting \ers{181221.1415} and \eqref{181221.1413} 
into the Lagrangian in \er{181201.2037}, 
we obtain the following dual Lagrangian:
\begin{equation}
\begin{split}
{\cal L}'_{B',{\rm SUSY}}
&= 
-\fr{1}{2e} \int d^4 \theta 
(\Psi + \xi) \log \(\fr{\Psi + \xi}{e M^2}\)
+\fr{1}{4}\int d^2\theta \Upsilon^\alpha \Upsilon_\alpha
\\
&
\quad
+\fr{\xi}{2e}\int d^4\theta  \log |\Phi_0|^2
-\fr{1}{2ie}\int d^2\theta \Upsilon^\alpha J_\alpha
+ \hc
\end{split}
\label{181201.2051}  
\end{equation}
The Lagrangian is now given by the chiral superfields 
$\Upsilon_\alpha$ and $\Phi_0$.
The chiral superfield $\Upsilon_\alpha$ describes a
massive 2-form and its superpartners.
The first term is the kinetic term for the massive 2-form,
and the second term is the mass term.
The third and the fourth terms are
 the coupling between the massive 2-form superfield and 
the string current superfield.
These terms can be explicitly seen by the component expression 
of the Lagrangian in \er{181201.2051}.

In order to show the component Lagrangian, we 
define the component fields of the chiral spinor superfield 
$\Upsilon_\alpha$ as follows.
The $\theta = \b\theta = 0$ components are defined as
\begin{equation}
\lambda'_\alpha = + i  \Upsilon_\alpha|,
\quad  
\b\lambda'_{\d\alpha} = - i  \b\Upsilon_{\d\alpha}|.
 \end{equation}
The components given by first order spinor derivatives 
are
\begin{equation}
 \begin{split}
\bs{D}' &= 
-\fr{1}{4}(D^\alpha \Upsilon_\alpha 
+ \b{D}_{\d\alpha} \b\Upsilon^{\d\alpha})|,
\\
\sigma' &= 
\fr{1}{\sr{2}} \Psi |
= \fr{1}{2\sr{2}i}
(D^\alpha \Upsilon_\alpha - \b{D}_{\d\alpha}\b\Upsilon^{\d\alpha}) |,
\\
B'_{mn} &= 
-i
((\sigma_{mn})_\alpha{}^\beta D^\alpha \Upsilon_\beta
 - (\b\sigma_{mn})^{\d\alpha}{}_{\d\beta} 
\b{D}_{\d\alpha} \b\Upsilon^{\d\beta})|.
 \end{split}
\end{equation}
Here, $B'_{mn}$ is a (non-gauge) 2-form field.
The components defined by 
second order spinor derivatives are
\begin{equation}
 \begin{split}
&  \psi'_\alpha 
:= 
 \fr{1}{\sr{2}} D_\alpha \Psi| 
= +\fr{i}{4\sr{2}}D^2\Upsilon_\alpha|+ \der_{\alpha\d\beta}\b\Upsilon^{\d\beta}|, 
\\
&
\b\psi'^{\d\alpha} :=  
\fr{1}{\sr{2}} \b{D}^{\d\alpha}\Psi |
= -\fr{i}{4\sr{2}}\b{D}^2\b\Upsilon^{\d\alpha}|
- \der^{\d\alpha \beta}\Upsilon_\beta|. 
 \end{split}
\label{181224.1949}
\end{equation}
Since $\Upsilon_\alpha$ is a chiral superfield, 
the components of higher than the second order are given by 
spacetime derivatives of the lower components. 
For example, the exterior derivative on the 2-form field is 
expressed in terms of the superfield as follows:
\begin{equation}
\begin{split}
&H'_{mnp} :=
\der_m B'_{np} + \der_n B'_{pm} + \der_p B'_{mn}
= \fr{1}{4} \epsilon_{mnpq} 
(\b\sigma^q)^{\d\alpha\beta}[D_\beta,\b{D}_{\d\alpha}] \Psi|.
\end{split}
\end{equation}
By using these component fields, we obtain the 
component Lagrangian:
\begin{equation}
\begin{split}
{\cal L}'_{B',{\rm SUSY}}
&
=
-\fr{1}{2e(\sr{2} \sigma' + \xi)}
\(
(\der^m \sigma') (\der_m \sigma') 
-(*H')^m (*H')_m\)
\\
&\quad
-\fr{i}{2e(\sr{2} \sigma' + \xi)}
\(
\b{\psi}'_{\d\alpha} 
(\b\sigma^m)^{\d\alpha\alpha} \der_m \psi'_\alpha 
+
\psi'^\alpha  (\sigma^m)_{\alpha\d\alpha}\der_m \b{\psi}'^{\d\alpha}
\)
\\
&
\quad
- \fr{1}{2e(\sr{2} \sigma' + \xi)^2}
\psi'^\alpha (\b\sigma^m)_{\alpha\d\alpha} \b{\psi}'^{\d\alpha} 
 (*H')_m
-
\fr{1}{2e(\sr{2} \sigma' + \xi)^3} 
\psi'^\alpha\psi'_\alpha \b\psi'_{\d\alpha}\b\psi'^{\d\alpha}
\\
&\quad
-\fr{1}{16} B'^{mn}B'_{mn}
+\fr{1}{2}\bs{D}'^2 -\fr{1}{4}\sigma'^2
\\
&
\quad
-\fr{i}{2}
(
\lambda'^\beta (\sigma^m)_{\beta\d\beta} \der_m \b\lambda'^{\d\beta}
+
\b\lambda'_{\d\beta}
 (\b\sigma^m)^{\d\beta\beta} \der_m \lambda'_\beta
)
-\fr{1}{\sr{2}} i 
(\lambda'^\alpha \psi'_\alpha - \b\lambda'_{\d\alpha}\b\psi'^{\d\alpha})
\\
&
\quad
+\fr{1}{2\cdot 2! e} \tilde{J}^{mn}B'_{mn}
+\fr{1}{2e}(\sr{2} \sigma + \xi) J
\\
&\quad
-\fr{1}{\sr{2}e}
(j^\alpha \psi'_\alpha +\b{j}_{\d\alpha} \b\psi'^{\d\alpha}) 
+\fr{i}{2e}
(j^\alpha (\sigma^m)_{\alpha\d\beta} \der_m \b\lambda'^{\d\beta}
+
\b{j}_{\d\alpha} (\b\sigma^m)^{\d\alpha\beta} \der_m \lambda'_\beta
).
\end{split} 
\label{181230.2146}
\end{equation}
The term $B'^{mn}B'_{mn}$ is the mass term for the 2-form field.
The coupling between the 2-form field and the string current 
is represented by $\tilde{J}^{mn}B'_{mn}$.
In this Lagrangian, we find that there are 
the couplings between the fermionic component of 
the string current superfield $j_\alpha$ and
$\lambda'_\alpha $
 compared with the Lagrangian in \er{181214.1609}.
Note that the superpartners of the 1-form gauge field 
are also dualized to the chiral spinor superfield
in the Lagrangian in \er{181230.2146} due to SUSY
similarly to the Lagrangian in \er{181214.1609}.

\section{Summary}
\label{sum}
In this paper, we have derived the dual formulations of the SUSY Abelian Higgs model with the FI term in 4D.
In particular, we have focused on the dual transformations
of ANO vortex strings in ${\cal N} = 1$ superspace.
These formulations of the ANO vortex strings can be obtained by 
splitting the chiral superfield charged under the 
$U(1)$ gauge symmetry into the regular part and singular part.
For the regular part which does not have zero points,
we have dualized this part into a 2-form prepotential 
in a previously known way.
In both of the dual formulations, the superpartners of the phase of the scalar field and 1-form are dualized into 
the 2-form prepotential and chiral spinor superfield due to SUSY
in contrast to the bosonic case, respectively.

In the dual transformation to the system with the 2-form prepotential, 
we have shown that the singular part of the chiral superfield 
gives us the string current superfield 
which has singularities of the two-dimensional delta function.
The string current superfield is coupled with 
the 2-form prepotential or the chiral spinor superfield, 
and satisfies the current conservation law by the SUSY algebra.
This current conservation law is consistent with the gauge symmetry 
of the 2-form prepotential.
Furthermore, we have identified the components of the string current superfield.
There are vortex strings as well as their superpartners.
We have confirmed that the vortex strings in the 
string current superfield
are the same as the ones in the bosonic (non-SUSY)
Abelian Higgs model.

We have further dualized the Abelian Higgs model into 
a theory described by a massive 2-form field.
The dual transformation has also been obtained by 
the previously known way.
We have also shown that the string current superfield 
is coupled with the chiral spinor superfield into which
the massive 2-form field is embedded.

There are several future work.
One is the BPS conditions for the ANO vortex strings in the 
dual formulations.
We have not considered the BPS conditions for the 
ANO vortex strings, although the conditions are 
important in SUSY theories.
Thus, we should discuss the dualities 
of the BPS conditions on the ANO vortex strings.

Another is the dual formulations including superpotentials 
which uplifts flat directions of the D-term potential.
In particular, we may discuss 
the dual formulations of the so-called 
M-model~\cite{Gorsky:2007ip,Shifman:2007ce}, 
in which the D-term potential 
is uplifted by a superpotential with an additional 
neutral chiral superfield.

The physical meaning of the bosonic and the 
fermionic superpartners of the string current
should also be investigated.
These superpartners are defined by spinor derivatives of 
the singular part of the chiral superfield.
They are coupled with the superpartners of the 2-form prepotential
or the chiral spinor superfield of the massive 2-form.
It may be an open question whether such couplings 
are particular ones for SUSY theories or 
can be generalized to non-SUSY cases.

Mathematical structures of the string current superfield 
would be interesting.
In 4D ${\cal N} =1$ SUSY theories, 
the superspace expressions of closed or exact $p$-forms 
have been already known~\cite{Gates:1980ay}.
On the other hand, 
the string current superfield formulated in this paper
can be an example of a superspace extension of the 
closed 2-form $J_{mn}$ which cannot be expressed by an
exterior derivative of a globally well-defined 1-form.
The generalization of such properties of the string current 
superfield to other $p$-forms may 
be useful to discuss other topological solitons.

The SUSY Abelian Higgs model is the simplest 
Lagrangian consisting of a single vector superfield and a single chiral superfield. 
When such a theory is realized as a low-energy effective action, 
it usually contains higher derivative corrections. 
Ghost-free higher derivative terms 
for a vector superfield and chiral superfield are available 
in Ref.~\cite{Fujimori:2017kyi} and Refs.~\cite{Buchbinder:1994iw,Buchbinder:1994xq,Khoury:2010gb,Koehn:2012ar,Koehn:2012te,Adam:2013awa,Nitta:2014pwa,Nitta:2014fca,Nitta:2015uba,Nitta:2018yzb,Nitta:2018vyc}, respectively.
An extension of our duality with vortex strings 
in more general cases with higher derivative terms 
is one of future directions.

The dual transformations discussed in this paper 
can be extended to cosmic strings in 
SUGRA~\cite{Dvali:2003zh,Copeland:2003bj,Dvali:2003zj}.
It will be convenient to 
use conformal SUGRA~\cite{Cremmer:1982en,Kugo:1982cu,Kugo:1982mr,Kugo:1983mv,Butter:2009cp,Kugo:2016zzf,Kugo:2016lum}
when we discuss the dual transformations of ANO vortex strings,
since the canonically normalized Einstein--Hilbert term 
can be obtained by the superconformal gauge-fixing
without tedious super-Weyl rescalings~\cite{Kugo:1982mr}.
In particular, the conformal superspace 
formalism~\cite{Butter:2009cp} and $p$-form gauge theories in 
the conformal superspace~\cite{Aoki:2016rfz,Yokokura:2016xcf}
would be useful, since we can discuss dual transformations 
in a manifestly SUSY way.

One of important extension would be a non-Abelian extension. 
A $U(N)$ gauge theory coupled with $N \times N$ Higgs fields 
in the fundamental representation with common $U(1)$ charges
is known to admit a non-Abelian vortex accompanied with 
non-Abelian ${\mathbb C}P^{N-1}$ moduli 
\cite{Hanany:2003hp,Auzzi:2003fs,Eto:2004rz,Eto:2005yh,Gorsky:2004ad,Eto:2006cx,Eto:2006db}, 
see Refs.~\cite{Tong:2005un,Eto:2006pg,Shifman:2007ce,Shifman:2009zz} as a review.
A non-Abelian duality of a non-Abelian vortex in a non-SUSY case 
was done in the context of dense QCD 
\cite{Hirono:2010gq,Eto:2013hoa},
by using a non-Abelian 2-form field 
\cite{Seo:1979id,Freedman:1980us}.\footnote{
Instead of the full non-Abelian duality, 
a partial duality can be done by
focusing on Abelian diagonal components \cite{Hirono:2018fjr}.
}
There, a coupling between 
the ${\mathbb C}P^{N-1}$ fields localized on a vortex world-sheet 
and a non-Abelian 2-form field in the bulk
was obtained.
A non-Abelian duality of non-Abelian vortex strings in a SUSY case would be possible by a non-Abelian extension of a chiral spinor superfield including a non-Abelian 2-form field as a component \cite{Clark:1988gx,Furuta:2001kx}. 
Another possibility of extensions is   
the case of an $SU(2) \times U(1)$ gauge theory coupled with 
the triplet Higgs fields with an equal charge,
admitting a BPS Alice string~\cite{Chatterjee:2017jsi,Chatterjee:2017hya}.
This will be also possible by using a 
non-Abelian chiral spinor superfield.

It would be interesting to consider the dual transformations of ${\cal N}=2$ extended SUSY theories allowing ANO vortex strings as well as non-Abelian vortex strings~\cite{Hanany:2003hp,Auzzi:2003fs,Eto:2004rz,Eto:2005yh,Gorsky:2004ad,Eto:2006cx,Eto:2006db}. 
To this end, the framework discussed in Ref.~\cite{Kuzenko:2004tn} 
might be useful.
${\cal N}=2$ extended SUSY theories also admit several composite solitons containing vortices such as vortex strings ending on a 
domain wall~\cite{Gauntlett:2000de,Shifman:2002jm,Isozumi:2004vg}, 
a monopole confined by 
vortices~\cite{Shifman:2004dr,Hanany:2004ea,Eto:2004rz,Nitta:2010nd},
 Yang--Mills instantons trapped inside a 
vortex~\cite{Shifman:2004dr,Hanany:2004ea,Eto:2004rz},
 and intersecting vortex strings~\cite{Naganuma:2001pu,Eto:2005sw}. 
The dual transformations in the presence of these composite solitons would be one of interesting future directions. Along this line, a dual transformation of a vortex-monopole complex was already discussed in Ref.~\cite{Chatterjee:2014rqa}.

\subsection*{Acknowledgements}
R.~Y.~thanks Hiroshi Isono for helpful discussions.
This work is supported by the Ministry of Education,
Culture, Sports, Science (MEXT)-Supported Program for the Strategic Research Foundation at Private Universities `Topological Science' (Grant No.\ S1511006).
The work of M.~N.~is also supported in part by  
the Japan Society for the Promotion of Science
(JSPS) Grant-in-Aid for Scientific Research (KAKENHI Grant
No.~16H03984 and No.~18H01217).
The work of M.~N.~and R.~Y.~is 
also supported in part by a Grant-in-Aid for 
Scientific Research on Innovative Areas ``Topological Materials
Science'' (KAKENHI Grant No.~15H05855) 
and ``Discrete Geometric Analysis for Materials Design'' (KAKENHI Grant No.~17H06462) from the MEXT of Japan, respectively. 

\providecommand{\href}[2]{#2}
\end{document}